\documentstyle[sprocl]{article}

\input{psfig}

\def\Journal#1#2#3#4{{#1} {\bf #2}, #3 (#4)}

\def\PLB{{\em Phys. Lett.}  B}

\def\be{\begin{equation}}
\def\ee{\end{equation}}
\def\bea{\begin{eqnarray}}
\def\eea{\end{eqnarray}}

\begin{document}

\title{MECHANISMS FOR DIRECT BREAKUP REACTIONS
\footnote[2]{to appear in the proceedings of the RCNP-TMU SYMPOSIUM on Spins
in Nuclear and Hadronic Reactions, October 26--28 1999, to be published with
World Scientific Publishing Company}
}

\author{G. BAUR}

\address{Forschungszentrum J\"ulich, Institut f\"ur Kernphysik, 
 D-52425 J\"ulich, Germany  \\ E-mail: g.baur@fz-juelich.de}

\author{S. TYPEL and H. H. WOLTER}

\address{Sektion Physik, Universit\"at M\"unchen,
D-85748 Garching, Germany \\ E-mail: stefan.typel@physik.uni-muenchen.de,\\
hermann.wolter@physik.uni-muenchen.de}

\author{K. HENCKEN and D. TRAUTMANN}

\address{Institut f\"ur Theoretische Physik, Universit\"at Basel, \\
Klingelbergstra\ss{}e 82, CH-4056 Basel, Switzerland
\\ E-mail: k.hencken@unibas.ch, \\dirk.trautmann@unibas.ch}


\maketitle
\abstracts{We review some simple  mechanisms of breakup 
in nuclear reactions. We mention the spectator breakup, which is described in 
the post-form DWBA. The relation to other formulations is also indicated.
An especially important mechanism is Coulomb dissociation.
It is a distinct advantage that the perturbation due to the electric field 
of the nucleus is exactly known. Therefore firm conclusions can be drawn
from such measurements. 
Some new applications of Coulomb dissociation for nuclear astrophysics are 
discussed.
}

\section{Introduction}

In general, the dynamics of nuclear breakup reactions can be quite complicated.
We wish to discuss in this minireview style some limiting cases, which show 
some simple features.
One may regard the work of Oppenheimer and Phillips in 1935 \cite{O35,OP35} 
as a starting 
point of the present subject. They tried to explain the preponderance 
of (d,p)-reactions over (d,n)-reactions by a virtual breakup of the 
deuteron in the Coulomb
field of the nucleus before the actual nuclear interaction takes place.
Because of the Coulomb
repulsion of the proton this would explain the dominance of (d,p)-reactions.
In this context, Oppenheimer~\cite{O35} also treated the real breakup of the 
deuteron in the Coulomb
field of a nucleus. In the meantime, the subject has developed quite a lot.
In addition to the deuteron, many different kinds of projectiles
(ranging from light to heavy ions, including radioactive beams)
have been used at incident energies ranging from below the Coulomb barrier 
to medium up to relativistic energies. At higher energies,
simplifications arise in the theoretical description,
since one can use Glauber theory (or the sudden approximation
in a semiclassical framework).

In Chapter 2 we discuss the ``spectator breakup'' mechanism. 
The breakup occurs  due to the strong interaction of one of the constituents 
with the target nucleus, while the ``spectator'' moves on essentially 
undisturbed.  Since this subject has
been dealt with extensively in the past \cite{BRST84}, we wish to give a
very brief outline of the development over the last few decades,
providing some of the relevant 
references. (This is of course a biased view of the present authors.)
We find that the post-form DWBA is especially suited to treat these processes.
We discuss in Chapter~3 Coulomb breakup using this post-form DWBA formalism. 
This approach 
has beeen used for a long time for the breakup of the deuteron. Recently,
this formalism has also been applied to the breakup of other halo nuclei 
with a simple structure, like ${}^{11}$Be. Also,
using the adiabatic (in a sense to 
be explained below) approach, a formula reminiscent of the formula for the
post-form DWBA 
has recently been developed. The relation between the two formulations 
will be discussed.

In Chapter 4 we discuss Coulomb dissociation using the semiclassical framework.
First and higher
order electromagnetic effects are treated. ``Post-acceleration'' can be viewed 
as a higher order electromagnetic effect. New possibilities for Coulomb 
dissociation experiments, also in the context of nuclear astrophysics are 
discussed in Ch.~5. Conclusions and an outlook are given in Chapter 6.

\section{Direct Breakup, Post-Form DWBA}
We assume a three-body model where a target nucleus A interacts
with a projectile $a=b+x$. The Hamiltonian is given by
\be
  H=T+V_{Ax}+V_{Ab}+V_{bx}.
\ee
This Hamiltonian is decomposed  as 
\bea \label{eq:Hi}
 H=H_{i}+V_{i} \quad \mbox{with} \quad
 H_{i}=T+V_{Aa}+V_{bx} \quad\mbox{and}\quad  V_{i}=V_{Ax}+V_{Ab}-V_{Aa},
\eea
where an (optical model) interaction $V_{Aa}$ is introduced. 
Another decomposition,
which is relevant for the final state, is given by
\be \label{eq:Hf}
 H=H_{f}+V_{f} \quad \mbox{with} \quad 
 H_f=T+V_{Ax} + V_{Ab} \quad\mbox{and}\quad V_f=V_{bx}.
\ee
We now use the latter decomposition to write 
the T-matrixelement for the elastic breakup reaction 
\be
   A+a \rightarrow A+b+x
\ee
in the post-form of the DWBA as (see, e.g., eq.~10 of Ref.~\citelow{BRST84})
\bea \label{eq:Tmat}
  \lefteqn{T_{\vec{q}_{a} \rightarrow \vec{q}_{b} \vec{q}_{x}} 
   =} \\ \nonumber & & 
  \int \!\!\! \int
  d^{3}r_{bx} \: d^{3}R_{Aa} \: \chi^{(-)\ast}_{\vec{q}_{b}}
  (\vec{R}_{b-Ax}) \chi^{(-)\ast}_{\vec{q}_{x}} (\vec{r}_{Ax})
  V_{bx}(\vec{r}_{bx}) \phi_{a}(\vec{r}_{bx})\chi^{(+)}_{\vec{q}_{a}}
  (\vec{R}_{Aa}),
\eea
where the $\chi$'s are the scattering wave functions of $a$, $b$ and $x$ 
generated by
the appropriate optical potentials. This is quite a complicated expression. 
Eikonal methods can be 
used to simplify it. For an intermediate model see, 
e.g., Ref.~\citelow{Angela99}. It contains
some simple limits, like the Serber model: see, e.g., Ref.~\citelow{BRST84}. 
Of course,
in the distorted waves of Eq. (5) the interaction of the target with the ``participant''
as well as the ``spectator'' is included to all orders in general.

\begin{figure}[htb]
\begin{center}
\leavevmode\psfig{file=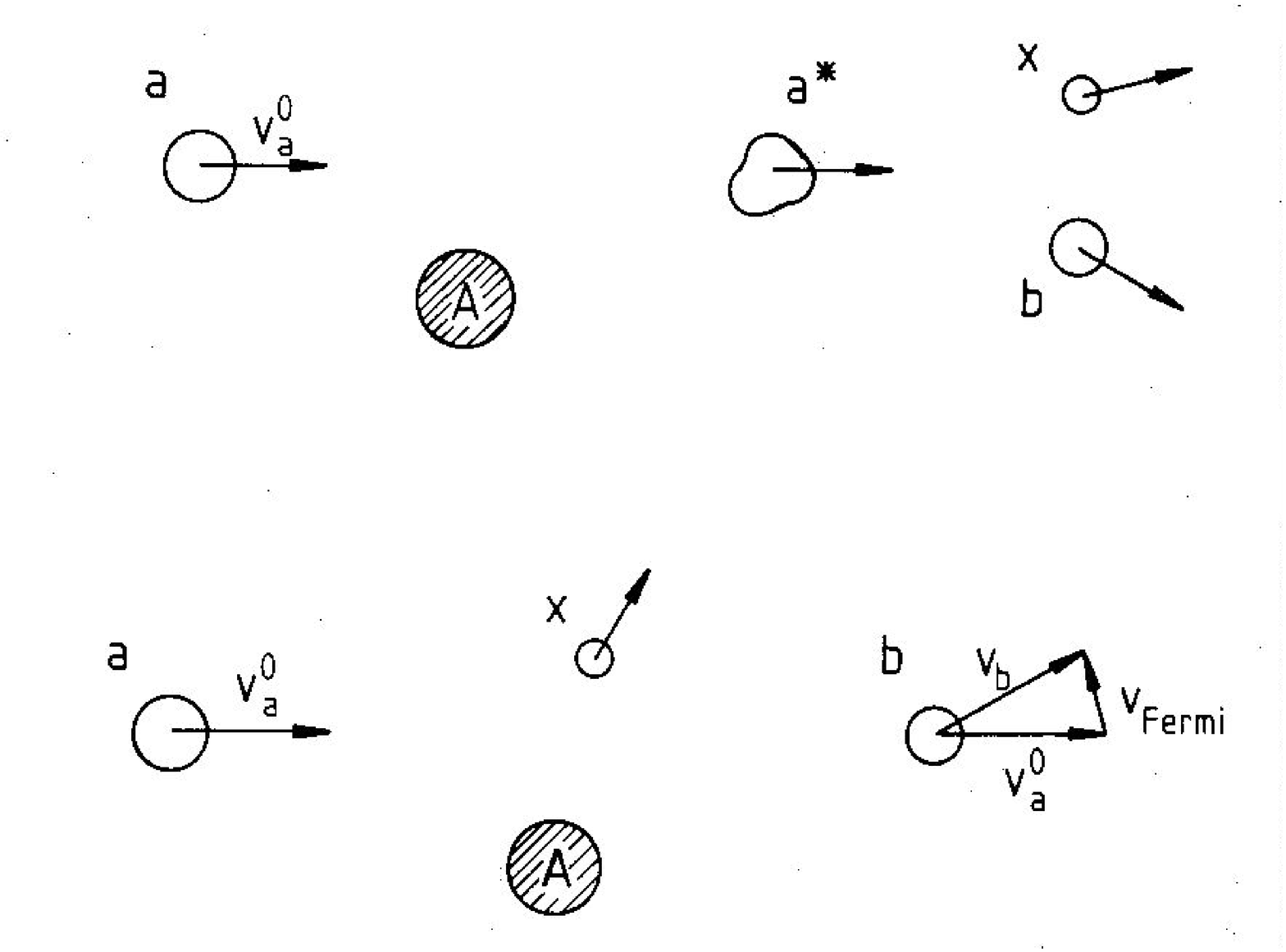,height=6.5cm}
\end{center}
\caption{Two basic reaction mechanisms for breakup are shown
schematically. In the upper figure, the projectile $a$ is excited to a
continuum (resonant) state $a^{\ast}$ which decays subsequently into the
fragments $b$ and $x$. In the lower part substructure $x$ interacts (in
all kinds of ways) with the target nucleus $A$, whereas $b=(a-x)$
misses the target nucleus (``spectator''). It keeps approximately the
velocity which it had before the collision. [Fig.~1 of
Ref.~\protect\cite{BRST84}.]}
\label{fig_prep1}
\end{figure}

An  expression identical to eq.~(\ref{eq:Tmat}) 
can also be written down in the prior form, but, as it stands
it would be very complicated to treat numerically, 
see, e.g., Ref.~\citelow{BT76}. We want to draw attention 
to a problem here which is related to the choice of the Hamiltionian in the
final state. If the breakup proceeds, e.g., through a resonance $a^{\ast}$ 
of the projectile
(resonance, or sequential type of breakup) the decomposition
eq.~(\ref{eq:Hi}) would certainly
be more appropriate to describe the final state
. The interaction $V_{bx}$ should be included in order to
properly describe that resonance. If the breakup proceeds directly into
the continuum, there seems to be no a priori reason to prefer one or the 
other decomposition. The two basic mechanisms are shown schematically
in Fig. \ref{fig_prep1}.  It was found that for Coulomb breakup of the 
deuteron at low energies, the choice Eq.~(\ref{eq:Hf}) is distinctly superior. 
In this way, the Coulomb interaction acts directly on the proton, and not on 
the centre of mass of the deuteron. For higher beam energies
it can be assumed that this distinction will become less and less important.
This problem is also reflected in the two choices in Ref.~\citelow{Mel99}.
Yet, this issue persists in general, and further clarification is necessary,
also in view of  ``post-acceleration'' to be discussed below. 
%
%

\begin{figure}[htb]
\begin{center}
\leavevmode\psfig{file=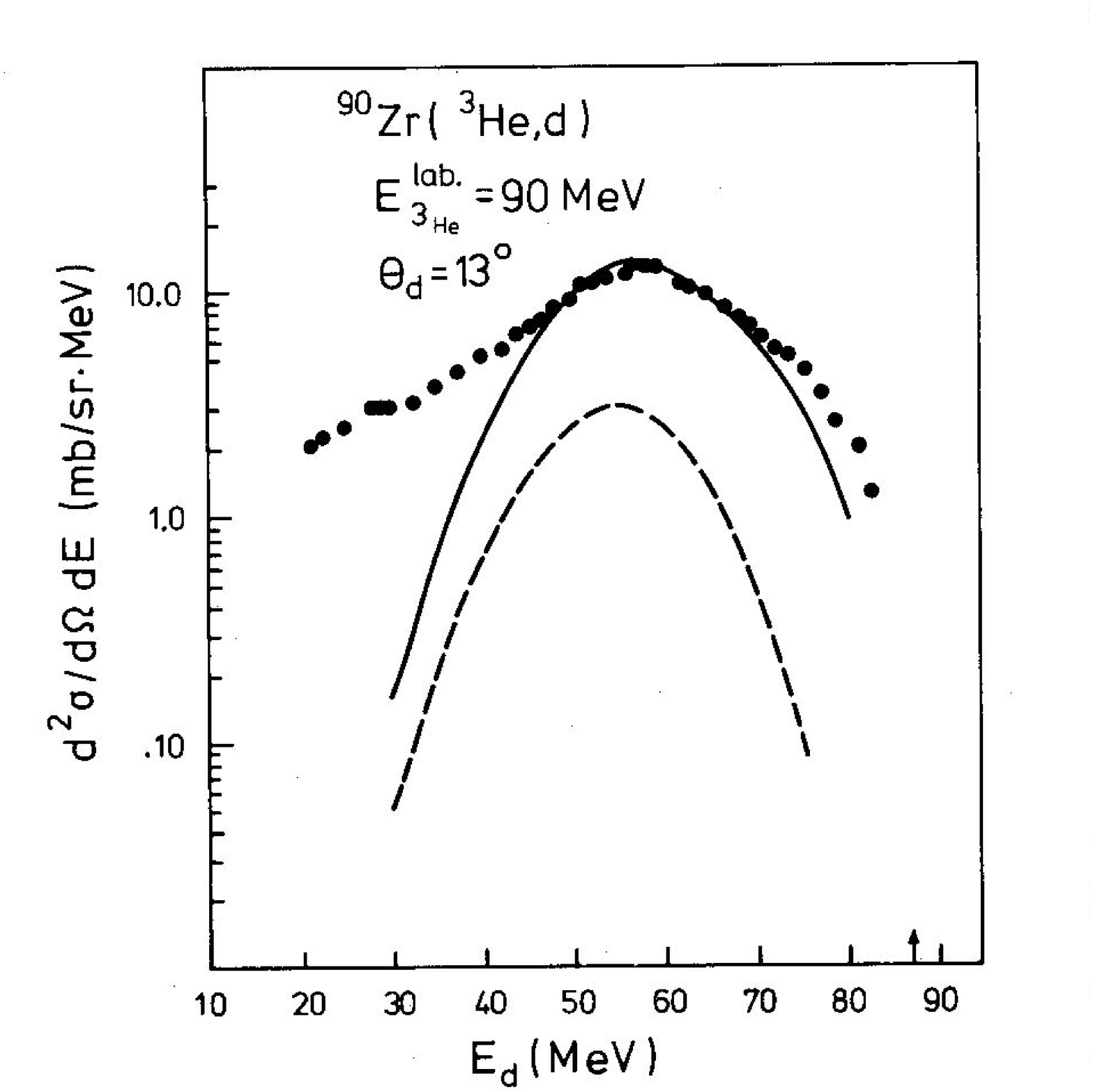,height=6.5cm}
\end{center}
\caption{Comparison of experimental ${}^{90}$Zr(${}^{3}$He,d) inclusive
spectra (Ref.~\protect\citelow{MSH80}) with theoretical calculations
\protect\cite{SBR80} (continuous line) for inclusive breakup. The
theoretically calculated contribution due to elastic breakup alone is
shown separately by the dashed line. The energy corresponding to the
three-body threshold is indicated by an arrow. [Fig.~11 of 
Ref.~\protect\cite{BRST84}.]}
\label{fig_prep11}
\end{figure}

It is of interest to treat also the case where the subsystem $B=x+A$ can 
go to other final channels $c$. This is especially simple when the 
``surface approximation''can be applied: Due to 
Coulomb repulsion and/or strong absorption  the ``wave function of 
the transferred particle''
\be
 \int d\xi_{A} \: \Phi^{(-)\ast}_{Bc} \Phi_{A} =
  4 \pi \sum_{l_{x}m_{x}} i^{l_{x}} F^{c}_{l_{x}}(r)
 Y_{l_{x}m_{x}}(\hat{r}) Y^{\ast}_{l_{x}m_{x}}(\hat{q}_{c})
\ee
has only to be known in the nuclear exterior. The integration in Eq. (6) is
over the $A$ nucleon variables. In this case the overlap 
integral is given in terms of the S-matrixelement of $x$-$A$ scattering,
which we denote by $S_{l_{x}c}$, as
\be
 F^{c}_{l_{x}}(r) = \delta_{l_{x}c} j_{l_{x}}(q_{x}r)
  + \sqrt{\frac{m_{x}q_{x}}{m_{c}q_{c}}} \frac{1}{2}
  \left( S_{l_{x}c}-\delta_{l_{x}c}\right)
  h^{(+)}_{l_{x}}(q_{x}r) \qquad (r\geq R_{0}) \: .
\ee
We neglect the spin of the particles, therefore the orbital angular 
momentum $l_{x}$
is the total angular momentum.
For charged particles $x$ the appropriate Coulomb functions have to be used
in place of the Hankel functions.
The validity of the surface approximation was checked by Kasano and 
Ichimura \cite{KI82}. It was found to be quite good for the (d,p) reaction at 
$E_{d}=26$~MeV. Inclusive breakup spectra were measured for many 
different systems and compared to theory. Agreement is generally
good \cite{BRST84}. Of course there are also pre-equilibrium and equilibrium 
contributions not considered in the inclusive breakup theory. As an example
we show in Fig.~\ref{fig_prep11} the ($^{3}$He,d) spectra measured in Osaka. 
%
%

The theory of inclusive breakup reactions was substantially generalized 
in a series 
of papers by M.~Ichimura, N.~Austern and C.~M.~Vincent (``IAV''). 
We give two references, from where
the full story can be traced back: \cite{IAV88,Ichi89}. In this series of  
papers, also
many formal aspects have been deeply elucidated and the relation of post-form
to prior-form DWBA (they give identical results) has also been made very clear.
Inelastic breakup reactions can also be used to study in an indirect way 
nuclear reactions
(much) below the Coulomb barrier (``trojan horse method''),
see Ref.~\citelow{BRST84,B86,B87,TW99}.
In the quasi-free approximation
the breakup cross section can be written as a product of
three factors: a) a kinematical factor, b) the momentum distribution
of particle $x$ in the projectile $a$ and c) an off-shell cross section
for the nuclear reaction to be investigated.
More general a relation between the S-matrix elements of the
two-body cross section and the actually measured three-body
cross section can be established with the help of the surface approximation
\cite{TW99}. 
In nuclear astrophysics, transfer reactions (like (d,p) or (${}^{3}$He,d),
or (Li,$\alpha$)) are used 
to study resonant states.
E.g., in the ${}^{22}$Na(${}^{3}$He,d)${}^{23}$Mg reaction 
states near the proton threshold were studied \cite{Ro95}. This is relevant 
for the hydrogen burning of ${}^{22}$Na.
In principle, also 
the continuum can be studied. E.g., the ``parallelism'' of (d,p) and 
(n,n)-reactions 
has been beautifully shown already in 1971, see Ref.~\citelow{Fu71}. 
The d+${}^{6}$Li reaction was investigated in Ref.~\citelow{Cher96} 
in this indirect way. Another recent
application is given in Ref.~\citelow{Claudio99} to the 
${}^{7}$Li(p,$\alpha$)${}^{4}$He-reaction.
An especially interesting case would be the indirect
study of the ${}^{12}$C($\alpha,\gamma$)${}^{16}$O 
reaction by means of a (${}^{7}$Li,t) or (${}^{6}$Li,d) reaction.
Quite recently \cite{Brun01}
the sub-Coulomb  $\alpha$-transfer reaction (${}^{6}$Li,d) and 
(${}^{7}$Li,t) to the bound
$2^{+}$ and $1^{-}$ states in ${}^{16}$O has been used to obtain 
information on the 
astrophysical S-factor.

\section{Spectator Mechanism for Coulomb breakup}

The Coulomb breakup (dissociation) of the deuteron can be viewed as 
spectator breakup, which was considered in the post-form DWBA in the previous
chapter: the Coulomb force acts solely on the proton and the neutron
moves on essentially undisturbed. Using a zero range approximation \cite{BT76},
the T-matrixelement separates into a ``bremsstrahlung-integral'' and 
a zero range
constant. The six-dimensional integral eq.~(\ref{eq:Tmat}) 
does not separate into two
three-dimensional integrals. In order to achieve this separation, 
one can use a ``local momentum approximation'' to pass from the 
coordinate $R_{Aa}$ (and $R_{b-Ax}$) to $R_{Ax}$
in the distorted wave $\chi_{\vec{q}_{a}}^{(+)}$. 
(The distorted wave $\chi_{\vec{q}_{b}}^{(-)}$ of
the neutron becomes a plane
wave.) For higher deuteron energies, this should be quite reasonable.
In its simplest version, one can use the asymptotic momentum 
$\vec{q}_{Aa}$ as 
a ``local momentum''. Now we have achieved a separation of the matrixelement
into the bremsstrahlungsintegral $T_{BS}$
and a matrixelement depending on the 
structure properties of the projectile
\be \label{eq:Tfac}
  T_{\vec{q}_{a} \rightarrow \vec{q}_{b} \vec{q}_{x}}
  \approx 
  T_{BS} (\vec{Q}) \:
  \int 
  d^{3}r_{bx} \: 
 \exp(-i\vec{P}\cdot \vec{r}_{bx}) 
  V_{bx}(\vec{r}_{bx}) \phi_{a}(\vec{r}_{bx})
\ee
with
\be
 \vec{Q} = \frac{m_{A}}{m_{A}+m_{x}} \vec{q}_{b} 
 \qquad \mbox{and} \qquad
 \vec{P} = \vec{q}_{b} -\frac{m_{b}}{m_{b}+m_{x}} \vec{q}_{a} \: .
\ee
Again, a Fourier transform of the ground state wave function of the
projeticle multiplied with the interaction potential is obtained.

\begin{figure}[htb]
\begin{center}
\leavevmode\psfig{file=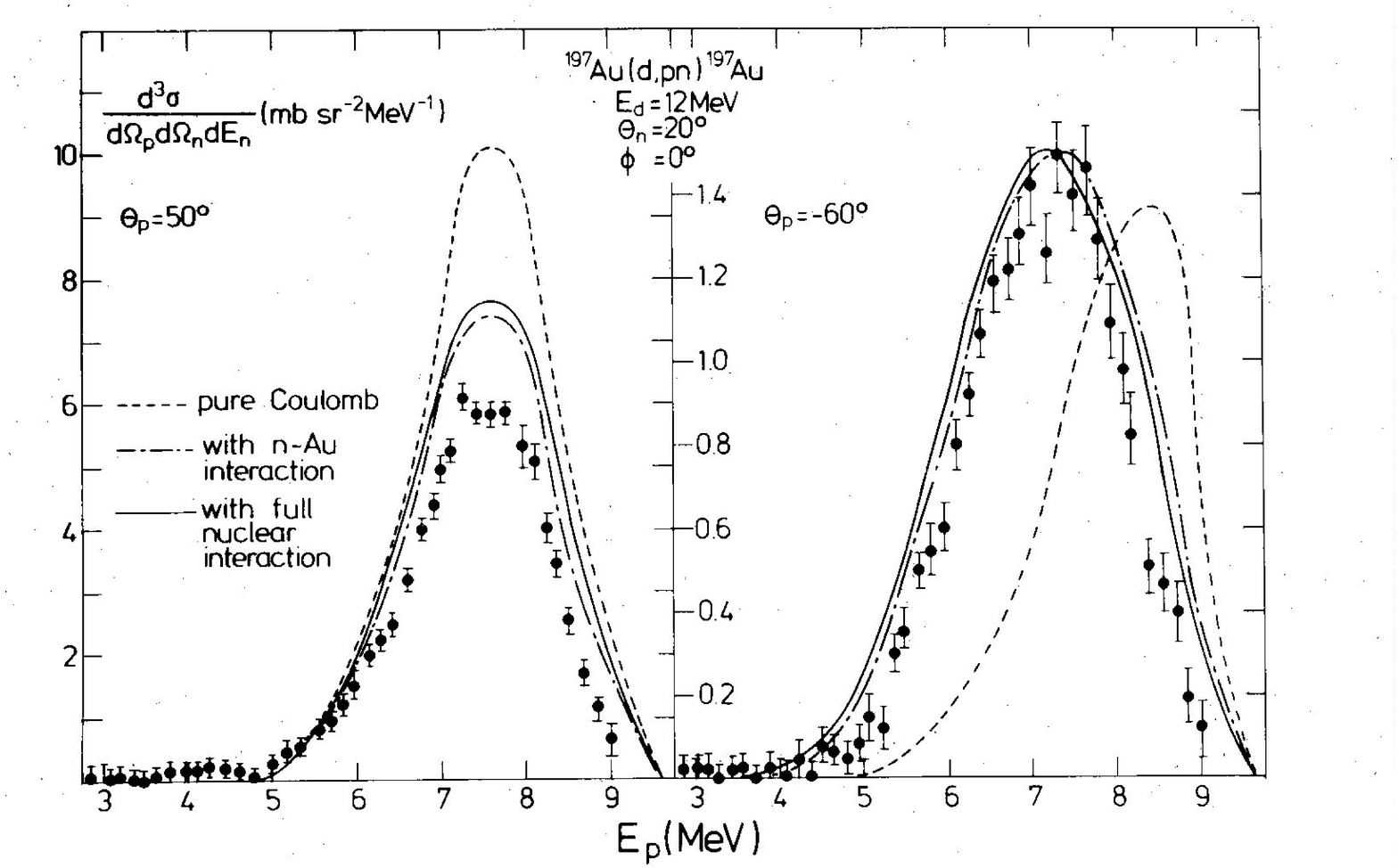,height=6.5cm}
\end{center}
\caption{Comparison of calculations \protect\cite{BT72} for the deuteron
breakup coincidence cross section on ${}^{197}$Au at $E_d= 12$~MeV with
the experimental results of Ref.~\protect\citelow{JLM73}. A coplanar geometry
was used, the neutron angle is $\theta_n=20^{\circ}$ and the proton angles
$\theta_p= 50^{\circ}$ and $\theta_p= - 60^{\circ}$ respectively. It can be
seen that, even at sub-Coulomb energies, the neutron-target
interaction strongly influences the breakup cross section, whereas the
$d+A$ and $p+A$ nuclear interactions in the initial and final channel
are very small. [Fig.~4 of Ref.~\protect\citelow{BRST84}.]}
\label{fig_prep4}
\end{figure}

Applications of this approach to reactions
below, around and somewhat above the Coulomb barrier have been extensively 
made, see, e.g., Ref.~\citelow{BRST84}. As an example we show in 
Fig.~\ref{fig_prep4} the p-n
coincidence spectra in Subcoulomb deuteron breakup on $^{197}$Au.
A strong ``postacceleration effect'' can be seen: the average energy
of the outgoing proton is much higher than the energy of the neutron.
This is due to the Coulomb repulsion. At higher deuteron energies,
this effect will be less pronounced.
This approach can also be extended to loosely bound neutron halo
nuclei, like ${}^{11}$Be \cite{SBB92}, or ${}^{19}$C . 
%
%

Coulomb dissociation of light nuclei
has recently been formulated in an ``adiabatic'' approach \cite{TRJ98,BTT98}:
It is assumed that 
projectile excitation is predominantly to states with low excitation energy.
One can say that the internal motion is slow as compared to the c.m.~motion 
of the projectile (``frozen nucleus approximation'' like in Glauber theory). 
In this approach, an adiabatic wave function is obtained. This wave function
is then inserted into an exact expression for the breakup amplitude. 
A matrix-element is obtained (eq.~21 of Ref.~\citelow{TRJ98}) 
which is identical to
the DWBA matrix-element (\ref{eq:Tfac}) given above. 
Numerical calculations for higher 
energy deuteron breakup are performed \cite{TRJ98} with good agreement with 
the data. We note that in this approach as well as in the DWBA method, the 
Coulomb interaction between $x$ and $A$ is treated non-perturbatively.
It is somewhat curious to see that the same formula is obtained
by using different approximation schemes.
In the following chapter, Coulomb dissociation
is treated in the initial state decomposition, eq.~(\ref{eq:Hi}).
These descriptions are
more closely connected to the one in this chapter
through the post-prior identities as one might 
first think. It should also be remembered that the 
approach in this chapter can only work when the neutron
is emitted directly into the continuum, with no further 
interaction with the ``core''. This works well for the deuteron 
which has no resonances in the continuum, 
and probably less well in other cases. 

\section{Theory of Electromagnetic Excitation and Dissociation}

\subsection{General}
It is interesting to study nuclear collisions where the colliding nuclei
interact only electromagnetically. This can be achieved
by using bombarding energies below the Coulomb barrier or by choosing
very forward scattering angles in high energy collisions. With increasing
beam energy states at higher energies can be excited; this can
lead, in addition to Coulomb excitation, also to Coulomb 
dissociation, for a review see, e.g., Ref.~\citelow{Ber01}. (Cross-sections for
such processes at the forthcoming relativistic heavy ion colliders
RHIC at Brookhaven and LHC(Pb-Pb) at CERN/Geneva are 
huge \cite{RHIC89,Hen02}.
A new field of studies is opened there, this is, however, beyond the
scope of the present review.)
Such experiments are also feasible with secondary (radioactive) beams.
The electromagnetic interaction which causes the dissociation is well known
and therefore there can be a clean interpretation of the experimental data. 
This is of
interest for nuclear structure and nuclear 
astrophysics \cite{Bau01,Ver01,Bau02}. 
Multiple electromagnetic excitation can also be important. We mention two
aspects: it is a means to excite new nuclear states, like the double
phonon giant dipole resonance \cite{Bau02}; 
but it can also be a correction to the
one-photon excitation \cite{Typ01,Typ02,Typ03}.
In the equivalent photon approximation the cross section for an
electromagnetic process is written as
\begin{equation}
 \sigma = \int \frac{d\omega}{\omega} \: n(\omega) \sigma_{\gamma}(\omega)
\end{equation}
where $\sigma_{\gamma}(\omega)$ denotes the corresponding cross section
for the photo-induced process and $n(\omega)$ is called the equivalent
photon number. For high enough beam energies it can be well approximated
by
\begin{equation}
 n(\omega) = \frac{2}{\pi} Z^{2} \alpha \ln \frac{\gamma v}{\omega R}
\end{equation}
where $R$ denotes some cut-off radius. More refined expressions, which
take the dependence on multipolarity, beam velocity or
Coulomb-deflection into account, are available in the 
literature \cite{Ber01,Typ02,Win01}.  
The theory of electromagnetic excitation is well developed, for
nonrelativistic as well as relativistic projectile velocities.
In the latter case, an analytical result for all multipolarities was
obtained in Ref.~\citelow{Win01}. The projectile motion was treated classically
in a straight-line approximation. Using Glauber theory, the projectile
motion can be treated quantally \cite{Ber01,Typ03,Ber02,Mue01,Bau03}. 
This gives rise to characteristic diffraction effects. 
The main effect is due to the strong absorption at impact parameters less 
than the sum of the two nuclear radii.

Nuclear excitation also has to be taken into account. The nuclear breakup
of halo nuclei was studied (among others, see e.g. Refs.
\cite{Ber92,Ban95,Bar96})
in Refs. \cite{Mue01,Hen01}. 
The nuclear interaction of course is less precisely known 
than the Coulomb interaction. In Ref.
\cite{Hen01} the nuclear breakup was studied
 using the eikonal approximation as well as the Glauber
multiple particle scattering theory. 
 No Coulomb interaction was included in this 
approach, as the main focus was on the breakup on light targets. In
Ref. \cite{Mue01} on the other hand, the combined effect of both nuclear 
and Coulomb excitation is studied. The nuclear contribution to the 
excitation is generally found to be small and has an angular dependence 
different from the electromagnetic one. This can be used to separate such 
effects from the electromagnetic excitation.

\subsection{Higher Order Effects}
\label{sec4.2}

Higher order effects can be considered in a coupled channels approach,
or using higher order perturbation theory. This involves a sum over
all intermediate states $n$ considered to be important.
Another approach is to integrate the time-dependent 
Schr\"{o}dinger equation directly for a given model Hamiltonian 
\cite{Mel99,Esb01,paris99,typwo99,ikebana}.

If the collision is sudden one can neglect the time
ordering in the usual perturbation approach. The interaction can be summed up 
to infinite order. In order to obtain the excitation cross section, one has to
calculate the matrix-element of this operator between the initial and final
state (i.e.~the intermediate states $n$ do
not appear explicitly). A related
approach was developed for small values of the adiabaticity parameter $\xi$
(the ratio between collision time and excitation time) in
Refs.~\citelow{Typ01,Typ02,Typ03}. In a simple zero range model 
for the neutron-core interaction, analytical results were obtained 
for $1^{st}$ and $2^{nd}$ order electromagnetic excitation \cite{Typ01}.
 These analytical results can
shed light on the higher order effects, which may be present in the Coulomb
dissociation of loosely bound neutron halo nuclei, 
like ${}^{11}$Be \cite{Nak02}
and ${}^{19}$C \cite{Nak03}. Higher order effects were recently 
discussed by J.~A.~Tostevin \cite{Tos99}.
We can use the sudden approximation (corresponding to $\xi$=0). Low $\xi$
corrections in $1^{st}$ and $2^{nd}$ order can be applied using the 
formulae given 
in Ref.~\citelow{Typ01}.
Higher order effects decrease strongly (like $1/b^4$) with impact parameter;
for grazing
collisions the $\xi$-value of the RIKEN experiments \cite{Nak02,Nak03} 
is of the order of 
0.1. We can use eq.~33 of Ref.~\citelow{Typ01}, where higher order effects are 
fully included.
We note the following points: for the angle integrated cross-section the 
$2^{nd}$ order
leads to an increase, whereas the interference of $1^{st}$ and 
$3^{rd}$ order leads to a 
decrease, an effect also found by J.~A.~Tostevin.
Of course, in the  expression  for the sudden approximation
all these higher order effects are included automatically.
It would be nice to have some numerical results, say for the ${}^{11}$Be
or ${}^{19}$C systems~\cite{Nak02,Nak03}. It can be supposed that they would
match closely to the results of Ref.~\citelow{Tos99}. This is because the same
physics is described, and higher order electromagnetic effects 
are included to all orders in both approaches. More work can certainly clarify
this issue.

\subsection{Some Results}
In this context it is also worthwhile to remember that electromagnetic
excitation of intermediate energy (exotic) beams has been recently
developed into a useful spectroscopic tool \cite{Mot01,Sch01}.
By measuring the excitation energies of the first $2^{+}$ 
states and the corresponding
B(E2)-values, nuclear structure effects like deformation,
can be studied in a 
unique way for nuclei far off stability. Electromagnetic excitation of 
the $1^{st}$ excited state 
in ${}^{11}$Be has been studied experimentally at GANIL \cite{Ann01}, 
RIKEN \cite{Nak01}
and MSU  \cite{Fau01}. This is a good test case, since the B(E1)-value of 
the corresponding ground-state transition is known already. Theoretical 
calculations \cite{Ber03,Typ04}
show that higher order effects are expected to be small.

Coulomb dissociation of exotic nuclei is a valuable tool to determine 
electromagnetic
matrix-elements between the ground state and the nuclear continuum. 
The excitation
energy spectrum of the ${}^{10}$Be+n system in the Coulomb dissociation
of the one-neutron
halo nucleus  ${}^{11}$Be on a Pb target at 72A MeV was measured \cite{Nak02}.
Low lying E1-strength 
was found. The Coulomb dissociation of the extremely neutron-rich nucleus 
${}^{19}$C was recently studied in a similar way \cite{Nak03}. 
The neutron separation 
energy of ${}^{19}$C could also be determined to be $530\pm130$~keV. 
Quite similarly, the Coulomb
dissociation of the 2n-halo nucleus ${}^{11}$Li was studied in various 
laboratories \cite{Kob89,Shi95,Zin97}. In an experiment at MSU \cite{Iek01}, 
the correlations of the
outgoing neutrons were studied. Within the limits of experimental accuracy,
no correlations were found.

In nuclear astrophysics, radiative capture reactions of the type
$  b + c \to a + \gamma $
play a very important role. They can also be studied in the time-reversed
reaction
 $ \gamma + a \to b + c \: $,
at least in those cases where the nucleus $a$ is in the ground state.
As a photon beam, we use the equivalent photon spectrum which is provided
in the fast peripheral collision. Reviews, both from an experimental
as well as theoretical point of view have been given  \cite{Bau01}, so we want
to concentrate here on a few points.

\begin{figure}[thb]
\begin{center}
\leavevmode\psfig{file=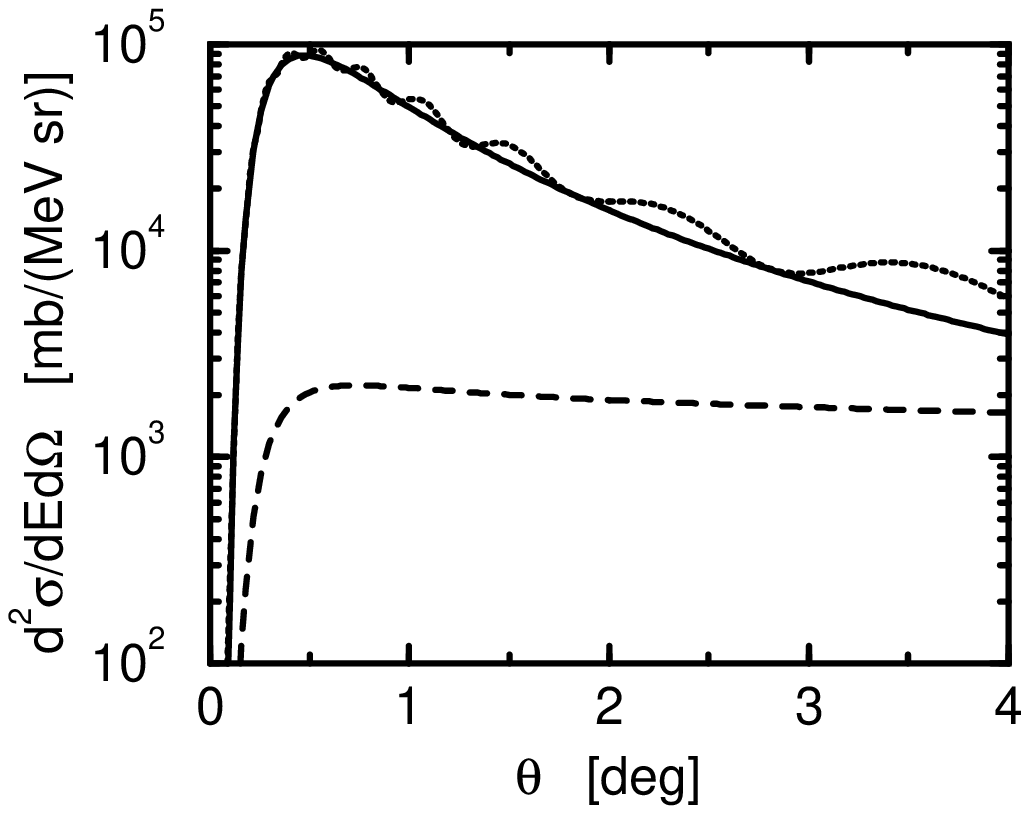,height=5cm,width=5cm}
\leavevmode\psfig{file=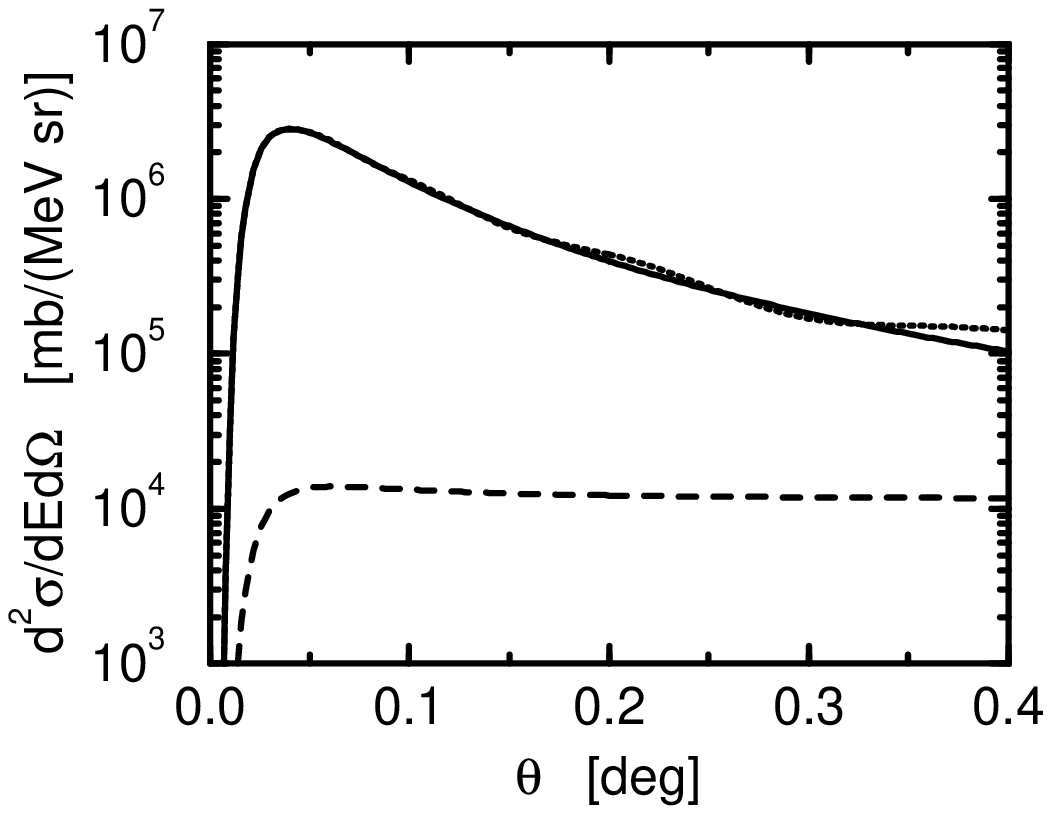,height=5cm,width=5cm}
\end{center}
\caption{Coulomb dissociation cross section of ${}^{8}$B scattered
on ${}^{208}$Pb as a function of the scattering angle for
projectile energies of 46.5~$A\cdot$MeV (left) and 250~$A\cdot$MeV
(right) and a ${}^{7}$Be-p relative energy of 0.3~MeV. 
First order results E1 (solid line), E2 (dashed line) and
E1+E2 excitation including nuclear diffraction (dotted line). 
[From figs.~4 and 5 of Ref.~\protect\cite{Typ03}.]} 
\label{procfig}
\end{figure}

The ${}^{6}$Li Coulomb dissociation into $\alpha$+d has been a test case 
of the method, see Ref.~\citelow{Bau01}. The ${}^{7}$Be(p,$\gamma$)${}^{8}$B 
radiative capture reaction is relevant for the solar neutrino problem. 
It determines the
production of ${}^{8}$B which leads to the emission of high energy neutrinos.
There are direct reaction measurements, for a recent one see 
Refs.~\citelow{Bog01,Ham98}.
Coulomb dissociation of ${}^{8}$B
has been studied at RIKEN \cite{Mot02}, MSU \cite{Kel01} and GSI 
\cite{Sue01,Iwa99}. 
Theoretical calculations are shown in Fig.~4. It is seen that E1 excitation
is large and peaked at very forward angles. E2 excitation is also there,
with a characteristically different angular distribution. Nuclear diffraction
effects are small. Altogether it is quite remarkable that completely 
different experimental methods with
different possible systematic errors lead to results that are quite consistent.

\section{Possible new applications of Coulomb dissociation 
for nuclear astrophysics}
Nucleosynthesis beyond the iron peak proceeds mainly by the r- and s-processes
(rapid and slow neutron capture) \cite{Rol01,Cow01}. 
To establish the quantitative details
of these processes, accurate energy-averaged neutron-capture cross
sections are needed. Such data provide information on the mechanism
of the neutron-capture process and time scales, as well as temperatures
involved in the process. The data should also shed light on neutron
sources, required neutron fluxes and possible sites of the
processes (see Ref.~\citelow{Rol01}). The dependence of direct neutron capture
on nuclear structure models was investigated in Ref.~\citelow{Rau98}. 
The investigated
models yield capture cross-sections sometimes differing by orders of magnitude.
This may also lead to differences in the predicted astrophysical r-process
paths. Because of low level densities, the compound nucleus model will not 
be applicable.

With the new radioactive beam facilities (either fragment separator or
ISOL-type facilities) some of the nuclei far off the valley of stability,
which are relevant for the r-process, can be produced. In order to assess
the r-process path, it is important to know the nuclear properties like
$\beta$-decay half-lifes and neutron binding energies. Sometimes, the
waiting point approximation \cite{Rol01,Cow01}
is introduced, which assumes an (n,$\gamma$)-
and ($\gamma$,n)-equilibrium in an isotopic chain. It is generally believed
that the waiting point approximation should be replaced by dynamic
r-process flow calculations, taking into account (n,$\gamma$), ($\gamma$,n)
and $\beta$-decay rates as well as time-varying temperature and neutron
density. In slow freeze-out scenarios, the knowledge of (n,$\gamma$)
cross sections is important.

In such a situation, the Coulomb dissociation can be a very useful tool
to obtain information on (n,$\gamma$)-reaction cross sections on
unstable nuclei, where direct measurements cannot be done. Of course,
one cannot and need not study the capture cross section on all the nuclei
involved; there will be some key reactions of nuclei close to
magic numbers. It was proposed \cite{Gai01} to use the Coulomb
dissociation method to obtain information about (n,$\gamma$) reaction
cross sections, using nuclei like ${}^{124}$Mo, ${}^{126}$Ru, ${}^{128}$Pd
and ${}^{130}$Cd as projectiles. The optimum choice of beam energy will
depend on the actual neutron binding energy. Since the flux of equivalent
photons has essentially an $\frac{1}{\omega}$ dependence, low neutron
thresholds are favourable for the Coulomb dissociation method. Note
that only information about the (n,$\gamma$) capture reaction to the
ground state is possible with the Coulomb dissociation method. The
situation is reminiscent of the loosely bound neutron-rich light nuclei,
like ${}^{11}$Be , ${}^{11}$Li and ${}^{19}C$.

In Ref.~\citelow{Typ01}
the $1^{st}$ and $2^{nd}$ order Coulomb excitation amplitudes
are given analytically in a zero range model for the neutron-core
interaction (see section~\ref{sec4.2}). 
This can be very useful to assess how far one can go down
in beam energy and still obtain meaningful results with the Coulomb
dissociation method. I.e., where the $1^{st}$ order amplitude
can still be extracted experimentally without being too much disturbed
by corrections due to higher orders. 
For future radioactive
beam facilities, like ISOL od SPIRAL, the maximum beam energy is an
important issue. We propose to use the handy formalism of Ref.~\citelow{Typ01}
to assess, how far one can go down in beam energy.
For Coulomb dissociation with two charged particles in the final
state, like in the ${}^{8}$B $\to$ ${}^{7}$Be + p experiment with
a 26~MeV ${}^{8}$B beam  \cite{vSc01} such simple formulae seem to be
unavailable and one should resort to the more involved approaches
mentioned in section \ref{sec4.2}.

A new field of application of the Coulomb dissociation method can be
two nucleon capture reactions. Evidently, they cannot be studied
in a direct way in the laboratory. Sometimes this is not necessary, where the
relevant information about resonances involved can be obtained by
other means (transfer reactions, etc.), like in the triple $\alpha$-process.

Two-neutron capture reactions in supernovae neutrino bubbles are studied
in Ref.~\citelow{Goe01}. 
In the case of a high neutron abundance, a sequence of two-neutron
capture reactions, ${}^{4}$He(2n,$\gamma$)${}^{6}$He(2n,$\gamma$)${}^{8}$He
can bridge the $A=5$ and 8 gaps. The ${}^{6}$He and ${}^{8}$He nuclei
may be formed preferentially by two-step resonant processes through their
broad $2^{+}$ first excited states  \cite{Goe01}. Dedicated Coulomb 
dissociation experiments can be useful,see \cite{au99}. Another key reaction can be the
${}^{4}$He($\alpha$n,$\gamma$) reaction \cite{Goe01}. 
The ${}^{9}$Be($\gamma$,n) reaction
has been studied directly (see Ref.~\citelow{Ajz01,hironeu}) 
and the low energy $s\frac{1}{2}$
resonance is clearly established. Despite this, a ${}^{9}$Be Coulomb
dissociation experiment could be rewarding (cf. also Ref.~\citelow{Kal01}).
Other useful information is obtained from (e,e') and (p,p') reactions on
${}^{9}$Be  \cite{Kue01}.

In the rp-process, two-proton capture reactions can bridge the waiting 
points \cite{Bar01,Goe02,Sch02}. From the ${}^{15}$O(2p,$\gamma$)${}^{17}$Ne, 
${}^{18}$Ne(2p,$\gamma$)${}^{20}$Mg and ${}^{38}$Ca(2p,$\gamma$)${}^{40}$Ti
reactions considered in Ref.~\citelow{Goe02}, 
the latter can act as an efficient reaction
link at conditions typical for X-ray bursts on neutron stars.
A ${}^{40}$Ti $\to$ p + p + ${}^{38}$Ca Coulomb dissociation experiment
should be feasible. The decay with two protons is expected to be
sequential rather than correlated (``${}^{2}$He''-emission).
The relevant resonances are listed in Table~XII
of Ref.~\citelow{Goe02}.
In Ref.~\citelow{Sch02} it is found that in X-ray bursts 2p-capture reactions
accelerate the reaction flow into the $Z \geq 36$ region considerably.
In Table~1 of Ref.~\citelow{Sch02} nuclei on which 2p-capture reactions may occur,
are listed; the final nuclei are ${}^{68}$Se, ${}^{72}$Kr, ${}^{76}$Sr,
${}^{80}$Zr, ${}^{84}$Mo, ${}^{88}$Ru, ${}^{92}$Pd and ${}^{96}$Cd
(see also Fig.~8 of Ref.~\citelow{Bar01}). It is proposed to study the Coulomb
dissociation of these nuclei in order to obtain more direct insight
into the 2p-capture process.

\section{Conclusion}
We discussed essentially two types of breakup mechanisms. However, we saw 
that they are somehow related, although more work should be still done to 
clarify this 
in more detail. The spectator mechanism is also useful as an indirect
method to study astrophysically relevant reactions below the Coulomb 
barrier. This is much in the same way as transfer reactions have been
traditionally used to study spectroscopic factors.
Peripheral collision of medium and high energy nuclei (stable or
radioactive) passing each other at distances beyond nuclear contact
and thus dominated by electromagnetic interactions are important tools
of nuclear physics research. The intense source of quasi-real
(or equivalent) photons has opened a wide horizon of related problems
and new experimental possibilities especially for the present and forthcoming 
radioactive beam facilities
to investigate efficiently
photo-interactions with nuclei (single- and multiphoton excitations
and electromagnetic dissociation).

\section*{Acknowledgments}
We have enjoyed collaboration and discussions on the present topics with
very many people; too many to name them all. We especially wish to mention
C.~A.~Bertulani, H.~Rebel, F.~R\"{o}sel, and R.~Shyam.
One of us (G.B.) wishes  to thank his Japanese hosts for the kind 
invitation to 
the RCNP/TMU symposium, he is also very grateful for their
 generous financial support.

\end{document}